# Magnetic Coupling in Ferromagnetic Semiconductor GaMnAs/AlGaMnAs Bilayer Devices

Y. F. Cao, Yanyong Li ,Yuanyuan Li, G. N. Wei, Y. Ji and K. Y. Wang[*]

*SKLSM, Institute of Semiconductors, CAS, P. O. Box 912, 100083, Beijing, P. R. China*

We carefully investigated the ferromagnetic coupling in the as-grown and annealed ferromagnetic semiconductor GaMnAs/AlGaMnAs bilayer devices. We observed that the magnetic interaction between the two layers strongly affects the magnetoresistance of the GaMnAs layer with applying out of plane magnetic field. After low temperature annealing, the magnetic easy axis of the AlGaMnAs layer switches from out of plane into in-plane and the interlayer coupling efficiency is reduced from up to 0.6 to less than 0.4. However, the magnetic coupling penetration depth for the annealed device is twice that of the as-grown bilayer device.

**magnetic coupling, magnetoresistance, bilayer structure, annealing, anisotropic field**

[*] E-Mail: kywang@semi.ac.cn

# 1 Introduction

Ferromagnetic semiconductor GaMnAs has attracted great interest not only because of its potential application in spintronics, but also the very rich fundamental physics [1-7]. The magnetic properties of GaMnAs are sensitive to Mn concentration, carrier density and also the strain, which allow us to tune the magnetic properties electrically and mechanically [3-7]. Giant Magnetoresistance (GMR) and Tunneling magnetoresistance (TMR) have been found in multilayer (Ga,Mn)As devices based on the principle of spin scattering transport and spin tunneling, which are potentially useful for information recording technology [8-12]. Exchange coupling is often used to pin the harder reference layer in metallic GMR and TMR systems. The interlayer coupling between the two GaMnAs layers can be modulated by changing the carrier concentration and the thickness of spacer layer, thus the spin transport properties also. Recently, the magnetic property of GaMnAs in GaMnAs bilayer films has been found to be strongly affected by the other ferromagnetic layer [13, 14] and the exchange bias between ferromagnetic metal MnAs and ferromagnetic semiconductor GaMnAs has been observed [15]. The overplayed Fe on top of the ferromagnetic semiconductor GaMnAs produces an above room temperature ferromagnetism at the interface of GaMnAs, which has been demonstrated by using XMCD [16]. However, the high conductivity of ferromagnetic metals prevents the direct magnetotransport measurements of the GaMnAs layer.

In order to understand the ferromagnetic coupling effect in GaMnAs bilayer system through magnetotransport measurements, we investigated the magnetic properties and the magnetotransport properties of the GaMnAs/AlGaMnAs bilayer. To investigate the magnetotransport properties of the bottom AlGaMnAs layer, we made the AlGaMnAs hall bar. Since the carriers in AlGaMnAs layer are strongly localized by the presence of the additional Al [17], it has extremely higher resistivity than the GaMnAs layer and also very different magnetic properties. With high This structure will allow us to investigate the magnetoresistance of the GaMnAs layer and understand the magnetic interaction between these two layers, which is hard to detect from direct magnetic property measurements. Compared the measured magnetic and magnetotransport results of the as-grown and annealed GaMnAs/AlGaMnAs bilayer devices, we find that the magnetic anisotropy of the bottom AlGaMnAs layer has strong effect on the magnetic coupling in this bilayer system, which could be very useful for designing spintronic devices in the future.

## 2 Experiment

The (Ga,Mn)As/(Al,Ga,Mn)As heterostructure was grown on semi-insulating GaAs (001) substrate by MBE at the temperature around 240 ℃. A 300 nm thick un-doped GaAs buffer layer was grown first, followed by 20 nm thick (Al,Ga,Mn)As and then 7 nm thick (Ga,Mn)As. The Al concentration is 30% and Mn concentrations are 6% for both layers. Circular devices with diameter 16 μm has been defined by using standard photolithography and wet etching for the bilayer structure [18]. For comparison, we also fabricated a [110] orientation Hall bar multi-terminal device with half etched down 9 nm to remove the top GaMnAs layer. There is a 10 um wide channel and 20 um spacing for 4 um wide bars. The metallic ohmic bond pads were evaporated with 50 nm Ti and 150 nm Au by electron beam evaporation. The magnetization properties of the bilayer have been measured by using super-conducting quantum interference device (SQUID). The magnetotransport properties of the devices were measured by using standard dc current methods with current I = 30 μA.

Remnant magnetizations along [110], [1$\bar{1}$0], and [001] orientations for the as-grown (Ga,Mn)As/(Al,Ga,Mn)As bilayer films were measured by using SQUID, which are shown in figure 1(a). A significant out of plane [001] magnetization component is observed below 4 K and almost unnoticeable at higher temperatures, which is attributed to the contribution of the as-grown bottom (Al,Ga,Mn)As layer with out-plane magnetic easy axis [19]. In contrast, much larger remnant magnetizations were observed along [110] and [1$\bar{1}$0] directions in the whole temperature range until up to the Curie temperature $T_C$ (45 K) of GaMnAs. It is attributed to the specific magnetic anisotropy of the top (Ga,Mn)As layer that the cubic anisotropy predominates at low temperatures and uniaxial anisotropy at higher temperatures as reported [20, 21], where the uniaxial and cubic anisotropy easy axes favor $[1\bar{1}0]$ and $<100>$ $[100]$, respectively. The carrier mediated ferromagnetism reorientation of easy axes is controlled by the anisotropy distribution of holes, which mainly depends on the Fermi level in valance band [22].

# 3 Results and discussion

The temperature dependent on the sheet resistance of AlGaMnAs/GaMnAs bilayer and bottom AlGaMnAs layer has been carefully investigated. The ratio of the sheet resistance of the device with only the bottom layer to that of the bilayer devices increases with decreasing temperatures, which is more than 70 below 65 K for the 24 hours fully annealed devices. This value is over 100 below this temperature for the as-grown devices, indicating the insignificant contribution of the magnetoresistance from the bottom layer in the bilayer device. The magnetoresistance (MR) versus perpendicular to the plane magnetic field for the as-grown bilayer devices at different temperatures are shown in figure 1b and 1c. At zero magnetic field, the resistance of ferromagnetic system is determined by its magnetic anisotropy and independent of the angle of the previously applied field for all measured temperatures [23]. Above the magnetic anisotropic fields marked as $B_{top}^{GaMnAs}$ in figure 1b, the resistance linearly decreases with increasing magnetic field further for all measured devices. This high field negative magnetoresistance has been attributed to suppression of weak localization and spin-disorder scattering at low and high temperatures, respectively [24, 25]. When the external magnetic field is smaller than the anisotropic magnetic field, very different behavior has been observed between Figure 1(b) and (c). At temperature (T = 20 K) well above the Curie temperature of bottom AlGaMnAs layer, the magnetoresistance with magnetic field below the anisotropic field is very similar to that of the single GaMnAs layer. The resistance of the bilayer device parabolically increases with increasing the perpendicular external magnetic field below the anisotropic field, which is due to the anisotropic spin-orbital interaction. It can be well fitted by a simple parabolic formula $R_{xx} \propto B^2$ as shown in Figure 1(c). The magnetic coupling from the bottom AlGaMnAs is expected to cause different magnetoresistance behavior of the single GaMnAs layer below the anisotropic magnetic fields when both layers are in ferromagnetic phase. Indeed, at T=3.5 K shown in Figure 1(b), the low magnetic field resistance did not simply increase parabolically with increasing the perpendicular external magnetic field. The resistance firstly increased parabolically from zero to $B_{bottom}^{GaMnAs}$, followed by another parabolic till $B_{top}^{GaMnAs}$ but with smaller coefficient. This is because the two parabolic magnetoresistances overlay in the magnetic field range, where the higher field second parabolic magnetoresistance also contributes to the lower field one. We then can write down the higher field second parabolic formula $R_{xx} = R_2 + \alpha_2 B^2$ and the low field first parabolic formula $R_{xx} = R_1 + (\alpha_1 + \alpha_2)B^2$, where $\alpha_1$ and $\alpha_2$ are the parabolic coefficients from the first and second parabolic magnetoresistance. The small hysteresis was observed in reverse side of the conjunction of two parabolas. This hysteresis is not originated from the misalignment of external magnetic field, since it does not disappear with carefully checking the misalignment of the perpendicular external magnetic field in ±10 degree. The hysteresis becomes weaker with increasing the annealing time as shown in Figure 2, indicating that the hysteresis is sensitive to the removal of the interstitial Mn. Surprisingly, at temperature T=13 K, which is above the Curie temperature of the as-grown AlGaMnAs layer, the magnetoresistance of the bilayer is still affected by the bottom layer. After subtracting the second parabolic component from the higher field, the low field resistance is linearly increased with external magnetic field. The physics origin of this linear field resistance will be discussed later in this paper. The behavior of the MR curves above 13 K is very similar to that of the single GaMnAs layer, where the typical MR measured at T=20 K is shown in Figure 1(c).

Using step annealing to remove the interstitial Mn, not only the conductance of GaMnAs is strongly enhanced, but also the magnetic anisotropy of GaMnAs has been modified. The magnetization easy axis of the bottom AlGaMnAs layer switches from out of plane to in-plane with increasing the annealing time, which will change the magnetic coupling between the two ferromagnetic layers. Thus the low field resistance of the bilayer GaMnAs/AlGaMnAs varies with different annealing time, which is shown in Figure 2 (a) and (b). Except for the device after 1 hour annealing, the positive magnetoresistance below $B_{top}^{GaMnAs}$ can be nicely fitted by combining two different parabolic curves, which is shown in the top of Figure 2(b). Both fitting parameters $\alpha_1$ and $\alpha_2$ decrease with increasing the annealing time. However, the low field resistance can not be fitted well with simple parabolic curve for the device after 1 hour annealing, which can be fitted well with adding an extra linear term. This is because the magnetization of device is close to the critical point where the magnetic easy axis switches from out of plane to in-plane. With increasing the annealing time further, the magnetization of the bottom layer is fully switched to in-plane. This is confirmed by the remnant magnetization along [110], [1$\bar{1}$0], and [001] orientations for 180 $^0$C annealed 48 hours (Ga,Mn)As/(Al,Ga,Mn)As bilayer films, which is shown in Figure 3(a). Compared with the as-grown films, there isn't any magnetic component for [001] orientation for the fully annealed film, which suggests the magnetization easy axes for the both layers are in the plane. The Curie temperature for the GaMnAs layer is about 100 K, and for the bottom layer is less than 40 K but higher than 20 K. The MR curve for the annealed (at 180 $^0$C for 24 hours) bilayer devices with magnetic field perpendicular to the plane at 20 K and 40 K are shown in Figure 3(b) and (c). The resistance dependent on the external magnetic field below anisotropic field can be well fitted by two simple parabolas at T=20 K. At T=40 K, a simple single parabolic like behavior is observed with perpendicular field below the anisotropic field, which is very similar to that of the single GaMnAs layer.

The magnetoresistance with applied perpendicular external magnetic field for the bilayer devices below the anisotropic field can be understood as follows: If there is not any magnetic coupling between the bottom AlGaMnAs layer and the top GaMnAs layer, the magnetoresistance of the bilayer device is expected to increase parabolically below the anisotropic field of the GaMnAs layer below the $T_C$ of GaMnAs. However, this is not consistent with the experimental observation. Our experimental results suggest that there is directly magnetic coupling between these two layers. The magnetic anisotropy of the bottom part of GaMnAs ($B_{bottom}^{GaMnAs}$) has been strongly suppressed by the AlGaMnAs layer, which results in a smaller magnetic field needed to pull the magnetization fully perpendicular to the plane compared to the top part of the GaMnAs. With magnetic field in the range between $B_{bottom}^{GaMnAs}$ and $B_{top}^{GaMnAs}$, the experimental resistance can be well fitted by a simple parabolic curve and without a linear term contribution. It suggests that the top part of the GaMnAs can strongly suppress the weak localization and spin-disorder scattering of the bottom part GaMnAs. After subtracting the parabolic contribution from the top part of the GaMnAs, the magnetoresistance change from the bottom part of the GaMnAs can be obtained. The ratio is the magnitude change of the resistance from the bottom part of GaMnAs to that of the whole GaMnAs layer. The ratio times the total thickness (7nm) of GaMnAs layer to obtain the magnetic coupling penetration depth (MCPD) into the GaMnAs layer from the bottom AlGaMnAs, which is shown in Figure 4(c). The MCPD is up to 4nm for the annealed device while less than 2nm for the as-grown device. The magnetic field needed to pull the magnetization of the bottom part of GaMnAs fully along z direction $B_{bottom}^{GaMnAs}$ can be directly obtained from the low field magnetoresistance, which is shown in Figure 4(a) for both the as-grown and the fully annealed devices. $B_{bottom}^{GaMnAs}$ for the as-grown device is smaller than that of the annealed devices. It should be noted that the value obtained through this method is invalid for the temperature above the Curie temperature of bottom AlGaMnAs layer. .When both layers of the bilayer device are in the ferromagnetic region, the ferromagnetic coupling efficiency $\beta$ between these two layers can be described as $\beta = \left(B_{top}^{GaMnAs} - B_{bottom}^{GaMnAs}\right)/B_{top}^{GaMnAs}$, which directly provides the ferromagnetic coupling strength. As shown in Figure 4(b), $\beta$ decreases with increasing the annealing time, which can be up to 0.6 for the as-grown device while less than 0.4 for the fully annealed device. With the increase of the annealing time, the main change of the

AlGaMnAs layer is that the Curie temperature increases and the easy axis switches from out of plane to in-plane. Compared with the magnetic coupling between the in-plane ferromagnetic materials with different magnetic anisotropy, the magnetic coupling between the perpendicular ferromagnetic materials and in-plane ferromagnetic materials has much stronger magnetic coupling efficiency but with much smaller MCPD.

# 4 Conclusion

In conclusion, we had carefully investigated the magnetic properties and magnetotransport properties of the (Ga,Mn)As/(Al,Ga,Mn)As bilayers. The magnetization easy axis for the top GaMnAs layer is always in the plane, while the magnetic easy axis of bottom AlGaMnAs layer switches from out of plane to in plane by low temperature annealing. We observed that the magnetic coupling between the two layers strongly affect the magnetotransport properties of the GaMnAs layer during the magnetization reversal with out of plane magnetic field applied. The magnetic anisotropy of the bottom layer switches from out of plane into in-plane by low temperature annealing. The interlayer coupling efficiency was found to be up to 0.6 for the as-grown device and less than 0.4 for the fully annealed bilayer device. However, the magnetic coupling penetration depth for the fully annealed device is twice that of the as-grown bilayer device. Our results show the possibility of tuning magnetoresistance and magnetic anisotropy with an added magnetic buffer layer having different types of relative magnetic easy axes between the bottom and top magnetic layers.

This work was supported by "973 Program" No. 2011CB922200, 2009CB929301, NSFC grant 11174272, 61225021 and EPSRC-NSFC joint grant 10911130232/A0402. KYW also acknowledges the support of Chinese Academy of Sciences "100 talent program".

## Figure Captions:

**Fig. 1:** (a) The temperature dependence of the projection of the remnant magnetization along different crystallographic orientations of the as-grown (Ga,Mn)As/(Al,Ga,Mn)As bilyaer film; (b) The resistance versus perpendicular external magnetic field for the as-grown (Ga,Mn)As/(Al,Ga,Mn)As bilayer device at (b) 3.5 K , (c)13 K and 20 K, where the red and green curves in figure b and c are fitting curves.

**Figure 2:** The resistance dependence of the external perpendicular magnetic field for the stepped annealed (Ga,Mn)As/(Al,Ga,Mn)As bilayer devices at 3.5 K (a) after 15 and 30 minutes annealing; (b) after 1 and 24 hours annealing, where the red solid and green dash dot curves are fitting curves. The blue curve in the top of figure 2b are fitting curve with an extra linear term.

Figure 3: (a) The temperature dependence of the projection of the remnan magnetization along different crystallographic orientations for 48 hour annealed (Ga,Mn)As/(Al,Ga,Mn)As bilayer film; The resistance versus perpendicular external magnetic field for 24 hour annealed (Ga,Mn)As/(Al,Ga,Mn)As bilayer device at (b) 20 K and (c)40 K, where the red solid and green dash dot curves in figure b and c are the fitting curves.

Figure 4: (a) The magnetic field needed to pull the magnetization of the bottom part of GaMnAs ($B_{bottom}^{GaMnAs}$) fully out of plane for the as-grown and 24 hour annealed bilayer devices; (b) the temperature dependence of the ferromagnetic coupling efficiency $\beta$ for both the as-grown and 24 hour annealed bilayer devices; (c) the temperature dependence of the magnetic coupling penetration depth (MCPD) for both the as-grown and 24 hour annealed bilayer devices.

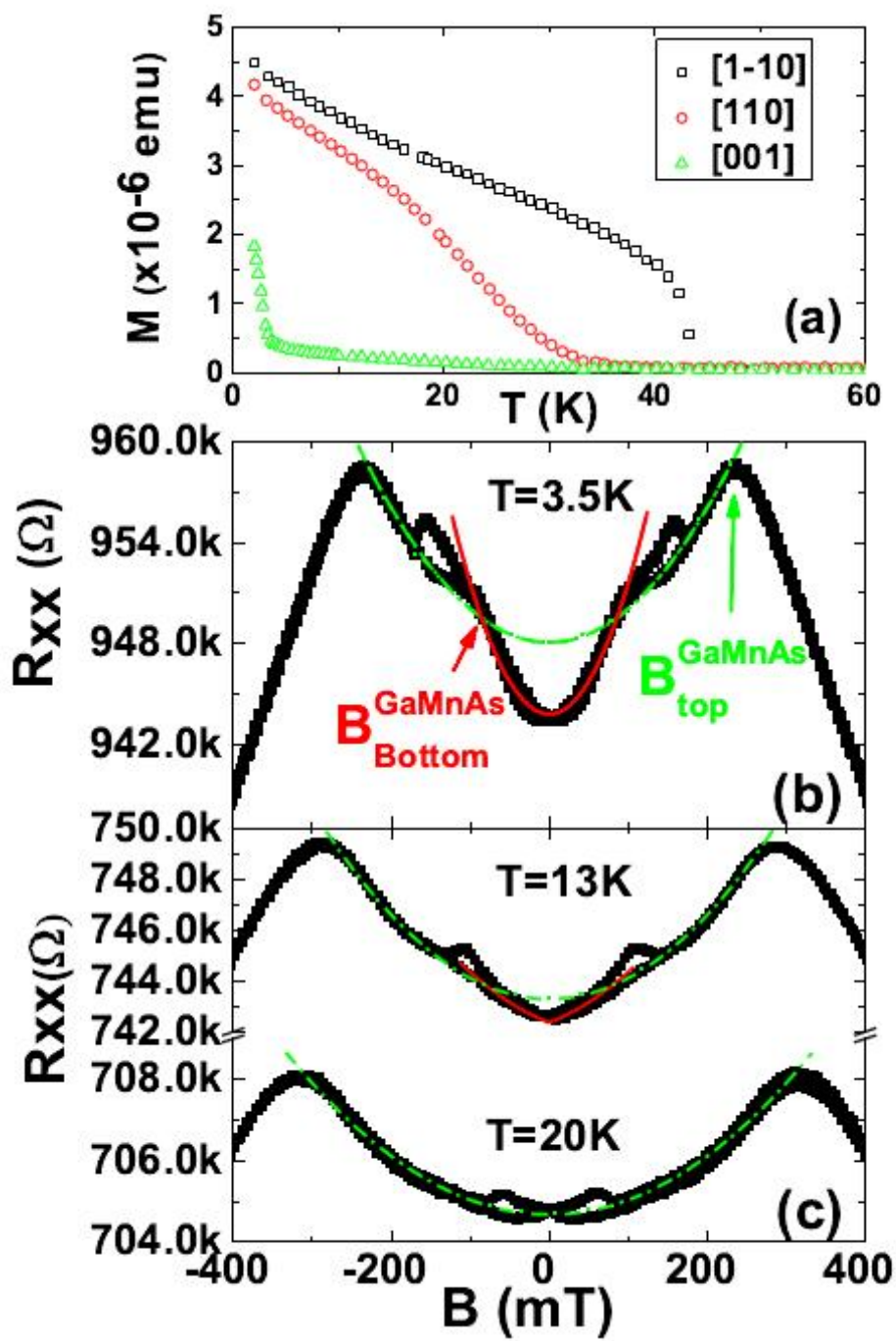

Figure 1 Y. F. Cao et al.

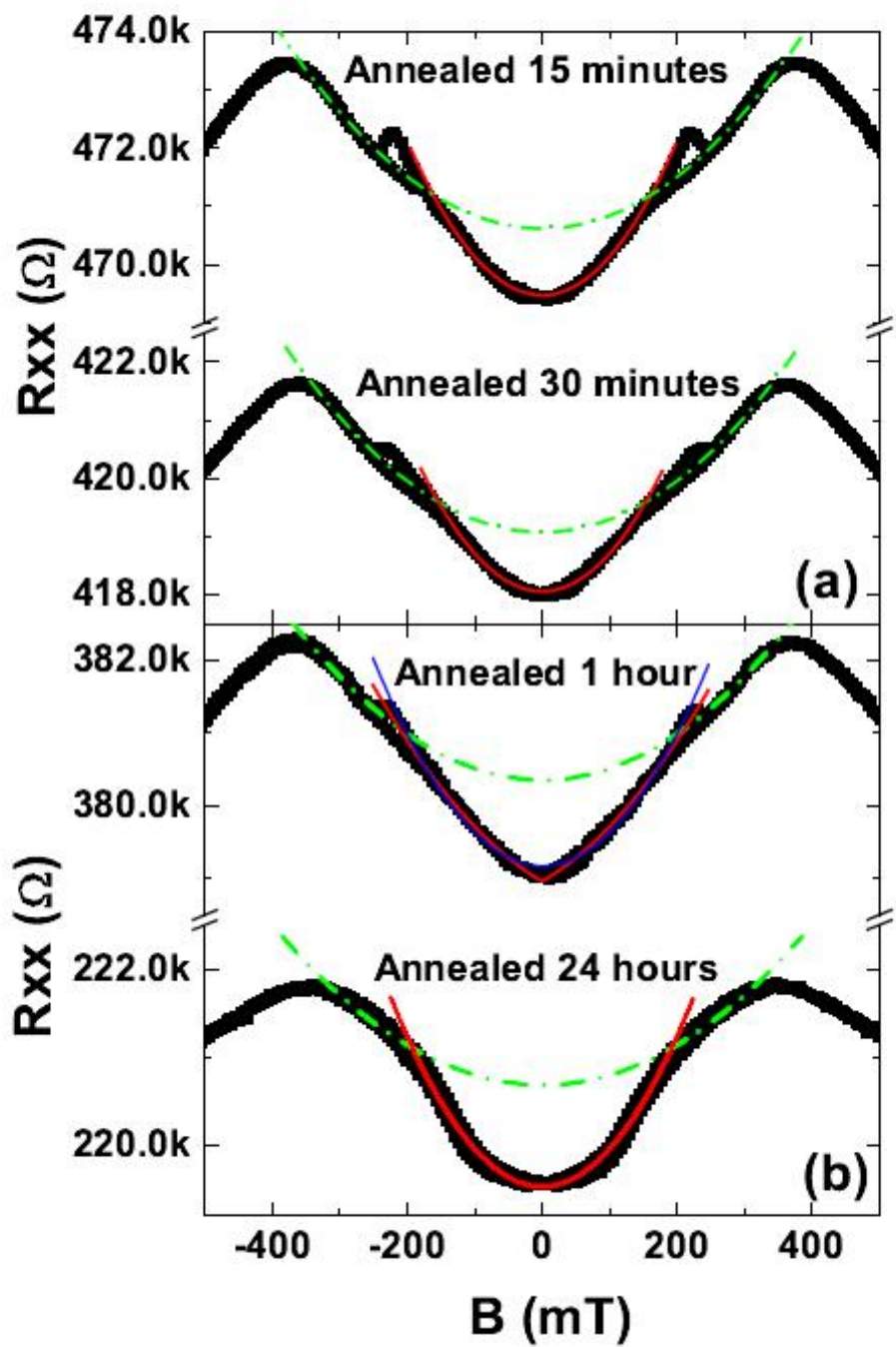

Figure 2 Y. F. Cao et al.

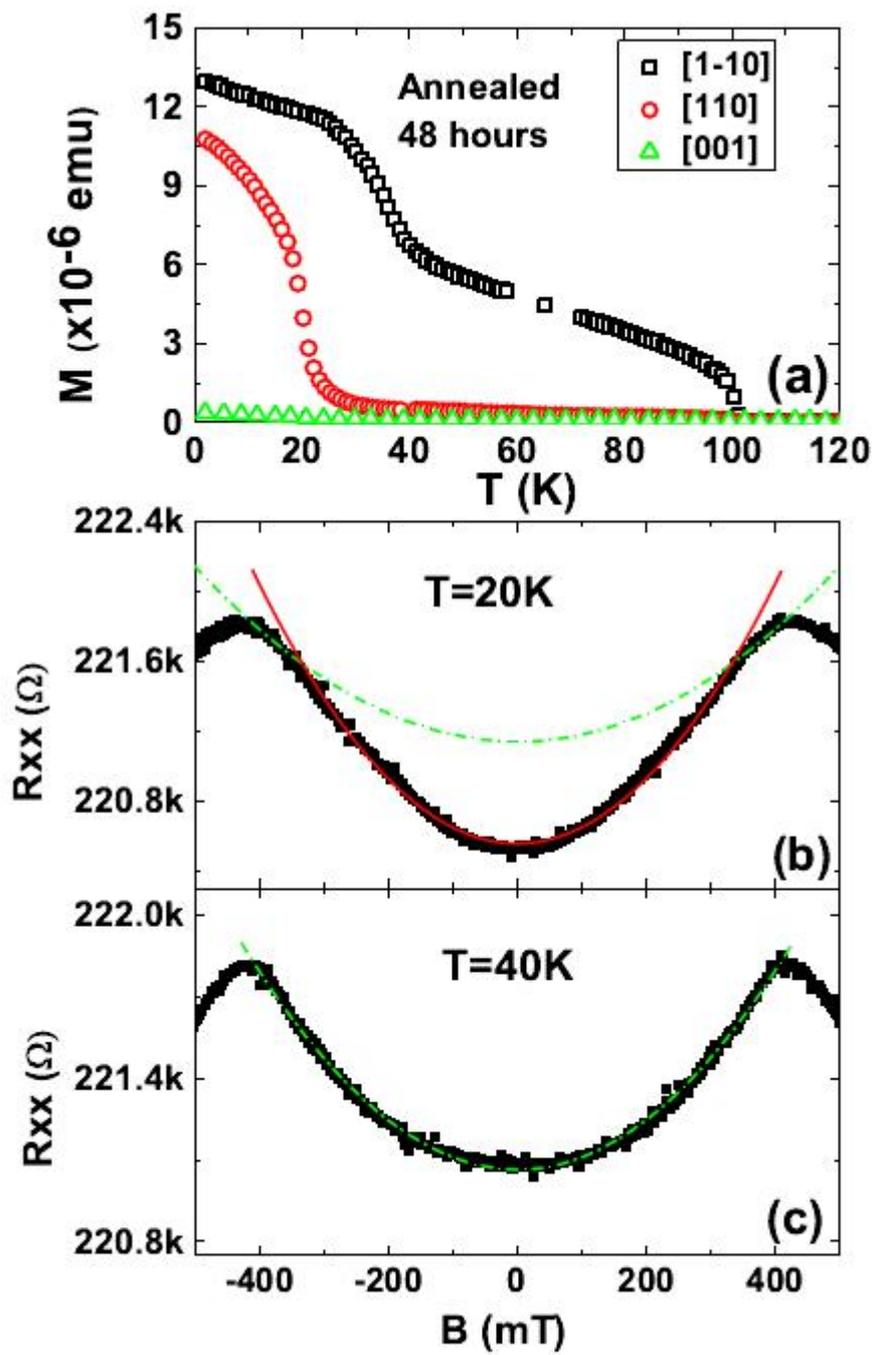

Figure 3 et al. Y. F. Cao et al.

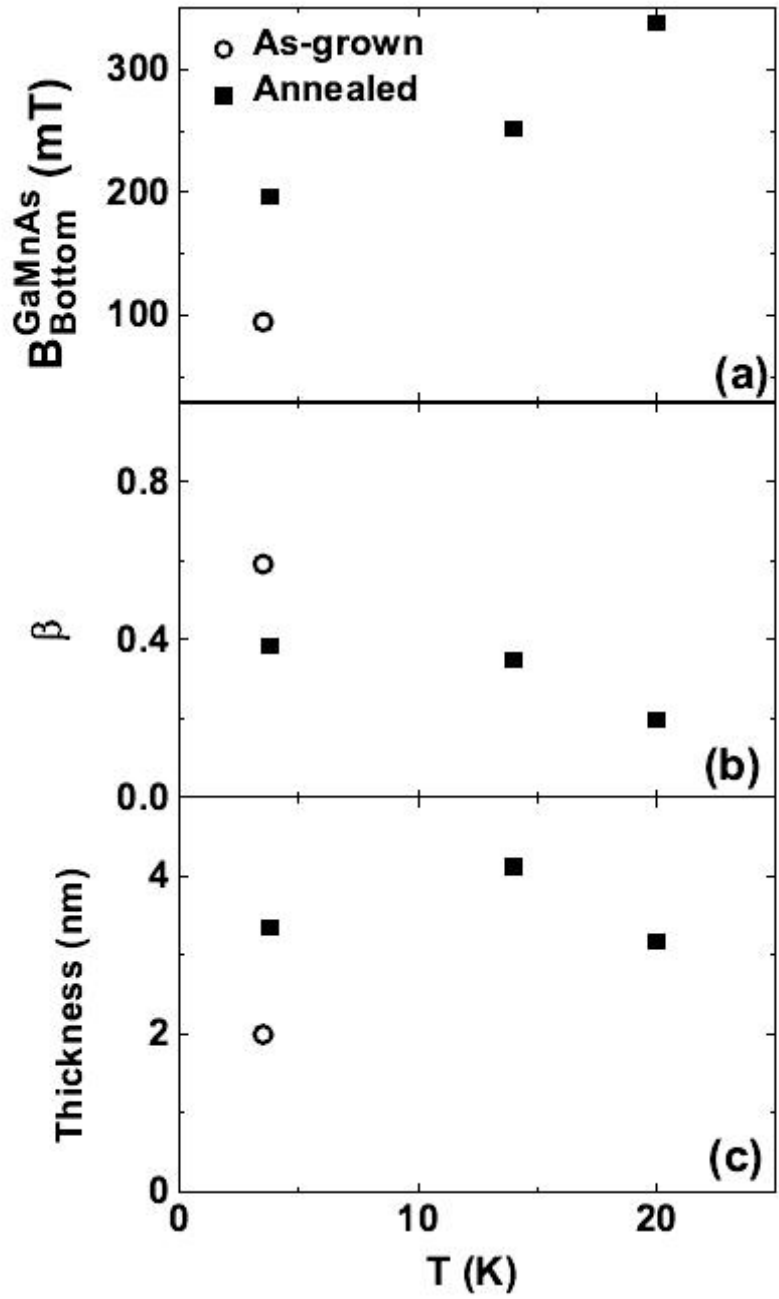

Fiugre 4. Y. F. Cao et al.